\documentclass[12pt,preprint]{emulateapj}

\begin{document}

\title{Lower-Luminosity Galaxies could reionize the Universe: Very
  Steep Faint-End Slopes to the $UV$ Luminosity Functions at
  $z\geq5$-8 from the HUDF09 WFC3/IR Observations\altaffilmark{1}}

\author{R. J. Bouwens\altaffilmark{2,3},
  G. D. Illingworth\altaffilmark{3}, P. A. Oesch\altaffilmark{3,9}, 
  M. Trenti\altaffilmark{4}, I. Labb\'{e}\altaffilmark{2,5}, 
  M. Franx\altaffilmark{2}, M. Stiavelli\altaffilmark{6}, 
  C. M. Carollo\altaffilmark{7}, P. van Dokkum\altaffilmark{8}, 
  D. Magee\altaffilmark{3}}
\altaffiltext{1}{Based on observations
  made with the NASA/ESA Hubble Space Telescope, which is operated by
  the Association of Universities for Research in Astronomy, Inc.,
  under NASA contract NAS 5-26555.}
\altaffiltext{2}{Leiden Observatory, Leiden University, NL-2300 RA
  Leiden, Netherlands} 
\altaffiltext{3}{UCO/Lick Observatory, University of California, Santa 
  Cruz, CA 95064}
\altaffiltext{4}{University of Colorado, Center for Astrophysics and
  Space Astronomy, 389-UCB, Boulder, CO 80309, USA}
\altaffiltext{5}{Carnegie Observatories, Pasadena, CA 91101}
\altaffiltext{6}{Space Telescope Science Institute, Baltimore, MD
  21218, United States}
\altaffiltext{7}{Institute for Astronomy, ETH
  Zurich, 8092 Zurich, Switzerland}
\altaffiltext{8}{Department of Astronomy, Yale University, New Haven, CT 06520}
\altaffiltext{9}{Hubble Fellow}
\begin{abstract}
The HUDF09 data are the deepest near-IR observations ever, reaching to
29.5$\,$mag.  Luminosity functions (LF) from these new HUDF09 data for
132 $z\sim7$ and $z\sim8$ galaxies are combined with new LFs for
$z\sim5$-6 galaxies and the earlier $z\sim4$ LF to reach to very faint
limits ($<0.05\,\,L_{z=3} ^{*}$). The faint-end slopes $\alpha$ are
steep: $-1.79\pm0.12$ ($z\sim5$), $-1.73\pm0.20$ ($z\sim6$),
$-2.01\pm0.21$ ($z\sim7$), and $-1.91\pm0.32$ ($z\sim8$).  Slopes
$\alpha\lesssim-2$ lead to formally divergent UV fluxes, though
galaxies are not expected to form below $\sim-10\,$AB mag.  These
results have important implications for reionization.  The weighted
mean slope at $z\sim6$-8 is $-$1.87$\pm$0.13.  For such steep slopes,
and a faint-end limit of $-10$ AB mag, galaxies provide a very large
UV ionizing photon flux.  While current results show that galaxies can
reionize the universe by $z\sim6$, matching the Thomson optical depths
is more challenging.  Extrapolating the current LF evolution to $z>8$,
taking $\alpha$ to be $-$1.87$\pm$0.13 (the mean value at $z\sim6$-8),
and adopting typical parameters, we derive Thomson optical depths of
$0.061_{-0.006}^{+0.009}$.  However, this result will change if the
faint-end slope $\alpha$ is not constant with redshift.  We test this
hypothesis and find a weak, though uncertain, trend to steeper slopes
at earlier times ($d\alpha/dz\sim-0.05\pm0.04$), that would increase
the Thomson optical depths to $0.079_{-0.017}^{+0.063}$, consistent
with recent WMAP estimates ($\tau=0.088\pm0.015$).  It may thus not be
necessary to resort to extreme assumptions about the escape fraction
or clumping factor.  Nevertheless, the uncertainties remain large.
Deeper WFC3/IR+ACS observations can further constrain the ionizing
flux from galaxies.
\end{abstract}

\section{Introduction}

One of the most important questions in observational cosmology regards
the reionization of the neutral hydrogen in the universe.  How did
reionization occur and which sources caused it?  Observationally, we
have constraints on reionization from the Gunn-Peterson trough in
luminous high-redshift quasars (Fan et al.\ 2002), the Thomson optical
depths observed in the Microwave background radiation (Komatsu et
al.\ 2011), and the luminosity function and clustering properties of
Ly$\alpha$ emitters (e.g., Ouchi et al.\ 2010).  Reionization appears
to have begun at least as soon as $z\sim11$ (e.g., Komatsu et
al.\ 2011) and finished no later than $z\sim6$ (e.g., Fan et al. 2007;
Ouchi et al.\ 2010).

Due to the very low volume densities of QSOs at high redshift (e.g.,
Willott et al.\ 2010) and the lack of evidence for other ionizing
sources (e.g., self-annihilating dark matter), star-forming galaxies
represent the most obvious source of ionizing photons.  However, given
the low volume densities of galaxies at high luminosities, the central
question at present is whether reionization can be accomplished
through a substantial population of very faint galaxies.  Such a
population is expected in many theoretical models (e.g., Trenti et
al.\ 2010; Salvaterra et al.\ 2011) and could, in principle, be
identified directly through extraordinarily deep observations (beyond
what is practical) or indirectly through the measurement of the
faint-end slope of the luminosity function and extrapolation to
fainter limits.  

The availability of the recently-obtained and deepest-ever near-IR
WFC3/IR observations over the Hubble Ultra Deep Field (HUDF: Beckwith
et al.\ 2006) with the HUDF09 program (GO 11563: PI Illingworth)
provide us with our best opportunity yet to quantify the prevalence of
very low-luminosity galaxies and the faint-end slope.  These
observations reach to $\sim$29.4-29.8$\,$mag at $5\sigma$
($\sim$2.5$\,$mag beyond $L^*$ at $z\sim5$-8), 0.4$\,$mag deeper than
the WFC3/IR observations available from the first year of the HUDF09
program (Bouwens et al.\ 2010; Oesch et al.\ 2010a).

\begin{figure}
\epsscale{1.15}
\plotone{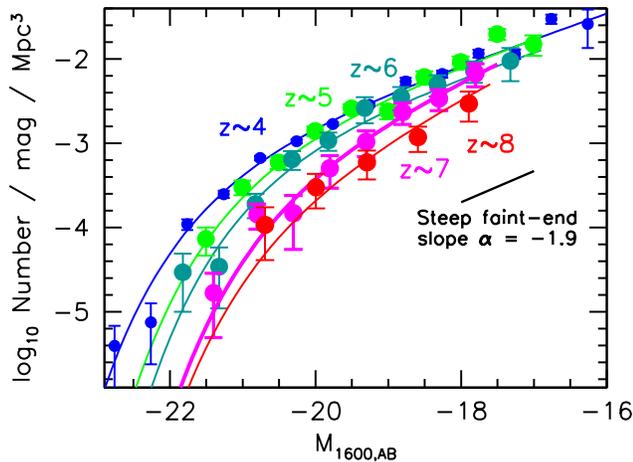}
\caption{The $UV$ luminosity functions at $z\sim4$, $z\sim5$,
  $z\sim6$, $z\sim7$ and $z\sim8$ (\S3).  The solid circles represent
  the stepwise maximum-likelihood determinations while the solid lines
  are the Schechter function determinations (they are \textit{not}
  fits to the points, though the overall agreement is excellent).  The
  $z\sim4$ constraints are from Bouwens et al.\ (2007).  The new
  $z\sim5$ (\textit{green}) and $z\sim6$ (\textit{cyan}) results are
  from the present work, while the $z\sim7$ (\textit{magenta}) and
  $z\sim8$ (\textit{red}) points are from Bouwens et
  al.\ (2011b).\label{fig:lf}}
\end{figure}

Here we take advantage of these new observations to quantify the
prevalence of extremely faint galaxies at $z\geq5$ and explore their
likely role in the reionization of the universe.  Our $UV$ LFs at
$z\sim5$-6 expand on the results of Bouwens et al.\ (2007) and take
advantage of the deep WFC3/IR observations.  These data reach
$\sim$0.5$\,$mag deeper than the existing $z_{850}$-band observations
over the HUDF and two HUDF05 fields.  The $z\sim7$-8 LFs we utilize
also rely on the full two-year HUDF09 data and were recently described
in Bouwens et al.\ (2011b).  We use these LFs to quantify how the UV
LFs and in particular the faint-end slope depends on cosmic time.  We
then calculate the flux in ionizing photons and the filling factor of
ionized hydrogen as a function of redshift and discuss the impact that
the faint galaxy population has on reionization.  The challenge will
be to match the large Thomson optical depths observed; recent results
(Oesch et al.\ 2009; see also Figure 3 here) have already shown that
observed galaxies can reionize the universe by $z\sim6$.  For ease of
comparison with previous studies, we adopt the concordance cosmology
$\Omega=0.3$, $\Omega_{\Lambda}=0.7$, and
$H_0=70\,$km$\,$s$^{-1}$Mpc$^{-3}$.  AB magnitudes (Oke \& Gunn 1983)
are adopted throughout.

\section{Observations}

Table~1 summarizes the search fields used for the $z\sim5$-8 LF
determinations and the approximate depths of the available ACS+WFC3/IR
observations.  Our primary data set consists of the full two-year
WFC3/IR observations of the HUDF and two flanking fields obtained with
the 192-orbit HUDF09 program (PI Illingworth: GO 11563).  Our second
data set is the $\sim$145$\,$arcmin$^2$ ACS+WFC3/IR observations over
the wide-area Early Release Science (Windhorst et al.\ 2011) and
CDF-South CANDELS (Grogin et al.\ 2011; Koekemoer et al.\ 2011)
observations.  We do not use the CANDELS observations for the
$z\sim7$-8 LFs since it is difficult to cleanly separate $z\sim7$ and
$z\sim8$ galaxies without $Y_{098}$/$Y_{105}$-band observations.  Such
observations are still lacking over most of the area.\footnote{While
  the $Y_{105}$-band observations were not available over the
  CDF-South CANDELS field when we derived the LFs described here, they
  are now available and have been used to improve the LF constraints
  at $z\sim8$ (e.g., Oesch et al.\ 2012b).}

\section{Results}

\subsection{Sample Selection and Possible Contamination}

The new $z\sim5$ and $z\sim6$ galaxy samples are identified to be
$\geq$5$\sigma$ sources in small $0.35''$-diameter apertures, after
first PSF-matching the ACS+WFC3/IR data to the WFC3/IR $H_{160}$-band
observations.  Our selection criteria for our $z\sim5$ sample is
identical to the $V_{606}$-dropout selection used by Bouwens et
al.\ (2007) and the $z\sim6$ selection uses the two-color criteria
$(i_{775}-z_{850}>1.3)\wedge(z_{850}-J_{125}<0.8)$ (also similar to
the Bouwens et al.\ 2007 criterion).  Sources must be undetected at
$>$2$\sigma$ in any band blueward of the Lyman break, at $>1.5\sigma$
in two such bands, or in the $\chi ^2$ image constructed from all data
blueward of the break (Bouwens et al.\ 2011b).

Applying these criteria to our five search fields (HUDF09, HUDF09-1,
HUDF09-2, ERS, and CDF-S CANDELS), we find 507 $z\sim5$
$V_{606}$-dropout galaxies and 203 $z\sim6$ $i_{775}$-dropout galaxies
(Table~1).

The only substantial source of contamination for our $z\sim5$-6
samples is from noise in our photometry and is estimated using the
same simulation techniques described in \S3.5.5 of Bouwens et
al.\ (2011b) for our $z\sim7$-8 samples.  These simulations suggest a
$\sim$10\% contamination rate overall, with less contamination
($\lesssim3$\%) at brighter magnitudes and more contamination
($\sim$10-20\%) at fainter magnitudes; we have corrected the observed
numbers accordingly.  Other possible sources of contamination include
low-mass stars, spurious sources, and transient sources (SNe), but we
estimate their contribution to be small ($<$1\%: see Bouwens et
al.\ 2011b).  For more details on the selection, we refer readers to
R.J. Bouwens et al.\ (2012, in prep).

\subsection{Luminosity Functions}

The present samples of $z\sim5$ and $z\sim6$ galaxies allow us to
extend current determinations of the $z\sim5$-6 $UV$ LF to even
fainter magnitudes.  In deriving the LFs, we utilize the same
maximum-likelihood procedures as described in Bouwens et al.\ (2007)
and Bouwens et al.\ (2011b: similar to those used by Sandage et
al.\ 1979 and Efstathiou et al.\ 1988).  As in Bouwens et al (2011b)
we derive the LFs by two approaches: (1) with a stepwise determination
that allows for a relatively model-independent determination of the
shape of the LF and (2) assuming the Schechter form and deriving the
best-fit parameters.

Selection volumes are estimated by adding model galaxies to the
observations and selecting them using the same procedure as with real
sources.  The model galaxy images are created by artificially
redshifting similar-luminosity $z\sim4$ $B_{435}$-dropout galaxies
from the HUDF (our deepest data set), scaling their sizes as
$(1+z)^{-1}$ to match the observed size-redshift scalings (e.g., Oesch
et al.\ 2010b).  We assume the $UV$ colors of the sources are
distributed as found by Bouwens et al.\ (2011c).

Our stepwise LFs and separately-established Schechter functions are
presented in Figure~\ref{fig:lf} as the solid points and dashed lines,
respectively.  Encouragingly, the Schechter form appears to be
generally well-matched to the stepwise determination (solid points),
with the possible exception of the $z\sim8$ LF where the stepwise LF
appears to be shallower.  However, as discussed in Bouwens et
al.\ (2011), this is due to differences in the way our two fitting
procedures cope with large-scale structure and two bright sources in
the HUDF09-2 field.  The best-fit Schechter parameters are provided in
Table~\ref{tab}.

The faint-end slopes $\alpha$ we determine for the $UV$ LF at $z\sim5$
and $z\sim6$ are $-1.79\pm0.12$ and $-1.73\pm0.20$, respectively,
consistent with our previous determination (e.g., Bouwens et
al.\ 2007) and a recent determination $-1.87\pm0.14$ at $z\sim6$ by Su
et al.\ (2011) given the large uncertainties.  Despite similar formal
uncertainties, the slopes are better determined, as they are less
impacted by systematics uncertainties from the use of (1) deeper data
and (2) a two-band LBG selection at $z\sim6$.  The quoted
uncertainties include the large-scale structure variance (20\% at
$z\sim5$-6: Trenti \& Stiavelli 2008; Robertson 2010a).

\begin{figure}
\epsscale{1.15}
\plotone{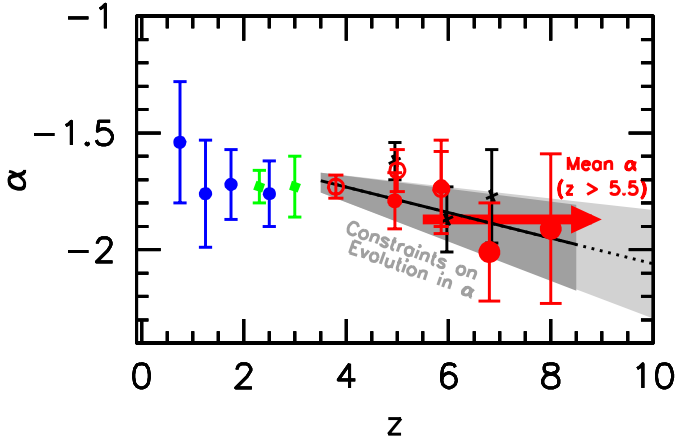}
\caption{Determinations of the faint-end slope $\alpha$ of the $UV$ LF
  versus redshift (\S3; \S4).  The large solid red points show the new
  slopes at $z\sim5$ and $z\sim6$ from this paper and those at
  $z\sim7$ and $z\sim8$ from Bouwens et al.\ (2011b).  Older
  determinations are the red open circles at $z\sim4$, $z\sim5$, and
  $z\sim6$ (Bouwens et al.\ 2007), black crosses at $z\sim5$, 6, and 7
  (Oesch et al.\ 2007; Su et al.\ 2011; Oesch et al.\ 2010), green
  squares at $z\sim2$-3 (Reddy \& Steidel 2009), and blue solid points
  at $z\sim0.7$-2.5 (Oesch et al.\ 2010c: see also Hathi et
  al.\ 2010).  Error bars are $1\sigma$.  The red horizontal line
  shows the mean faint-end slope $\alpha=-1.87\pm0.13$ we derive at
  $z\geq6$.  The solid black line is a fit of the $z\sim4$-8 faint-end slope
  determinations to a line, with the 1$\sigma$ errors (gray area:
  calculated by marginalizing over the likelihood for all slopes and
  intercepts).  The new WFC3/IR observations provide evidence that LFs
  at $z\geq5$-6 are very steep, with faint-end slopes
  $\alpha\lesssim-1.8$.
\label{fig:alpha}}
\end{figure}

\begin{figure}
\epsscale{1.20}
\plotone{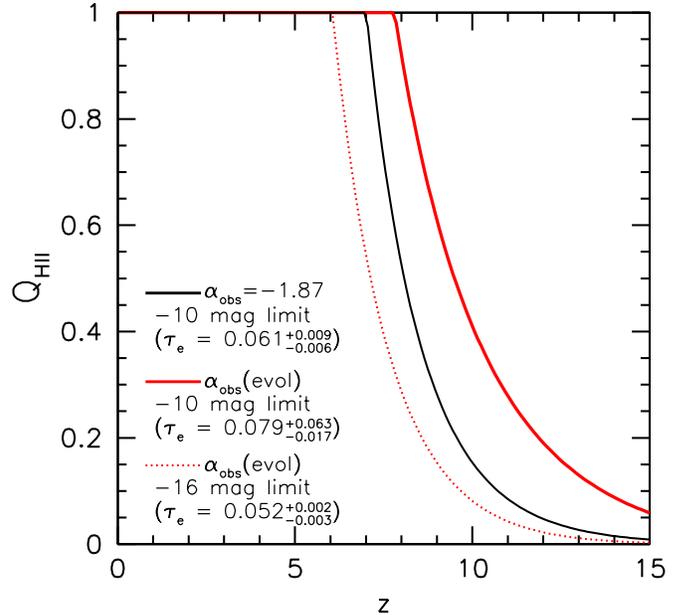}
\caption{Filling factor of ionized hydrogen $Q_{HII}$ versus redshift
  using our LF-fitting formula for UV LF at $z\geq4$ (Table~1).  The
  respective ionization histories (represented by the lines) were
  calculated from Eq.~2 assuming a Lyman-continuum escape fraction
  $f_{esc}$ of 20\%, a clumping factor of 3, an IGM temperature of
  $2\times10^4\,$K, a 1/50 $Z_{\odot}$ Salpeter IMF, and assuming the
  LF extends down to $-$10$\,$mag (with the same faint-end slope
  $\alpha$).  See text for references and see also Figure 8 from
  Bolton \& Haehnelt (2007) and Figure 4 from Oesch et al.  (2009).
  The solid black line shows the filling factor derived from best-fit
  LF (Figure~2 and Table~1: \S4) with the mean faint-end slope
  $\alpha=-1.87\pm0.13$ found at $z\sim6$-8 and for the best-fit
  evolution in $\alpha$.  The Thomson optical depths $\tau$ for these
  ionization histories are $0.061_{-0.006}^{+0.009}$ and
  $0.079_{-0.017}^{+0.063}$, respectively. The red dotted line is for
  our best-fit faint-end slope $\alpha$ evolutionary scenario, but
  assumes the LF extends to just $-$16$\,$mag (the limit of our data),
  showing that while observed galaxies can reionize the universe by
  $z\sim6$ they produce $\tau$'s, i.e., $0.052_{-0.003}^{+0.002}$,
  which are too low.  Changes in the adopted cosmology also affect the
  derived $\tau$.  Allowing for evolution of the faint-end slope and a
  faint-end limit of $-$10$\,$mag to the LF, we find an optical depth
  easily consistent with the $\tau=0.088\pm0.015$ found by WMAP
  (Komatsu et al. 2011).  This suggests that star-forming galaxies in
  the first 700-800 Myr could reionize the universe.\label{fig:reion}}
\end{figure}

\begin{figure*}
\plottwo{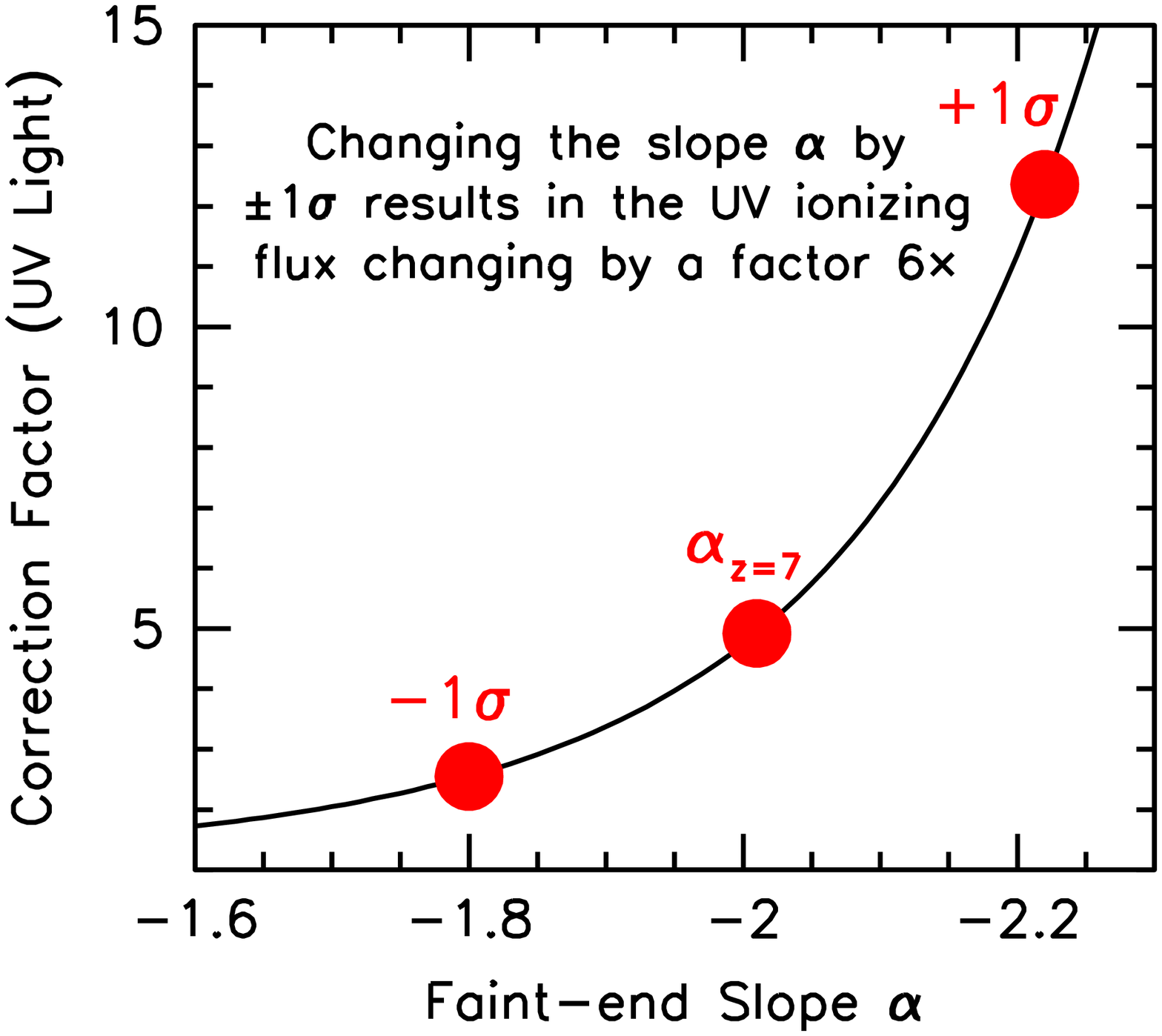}{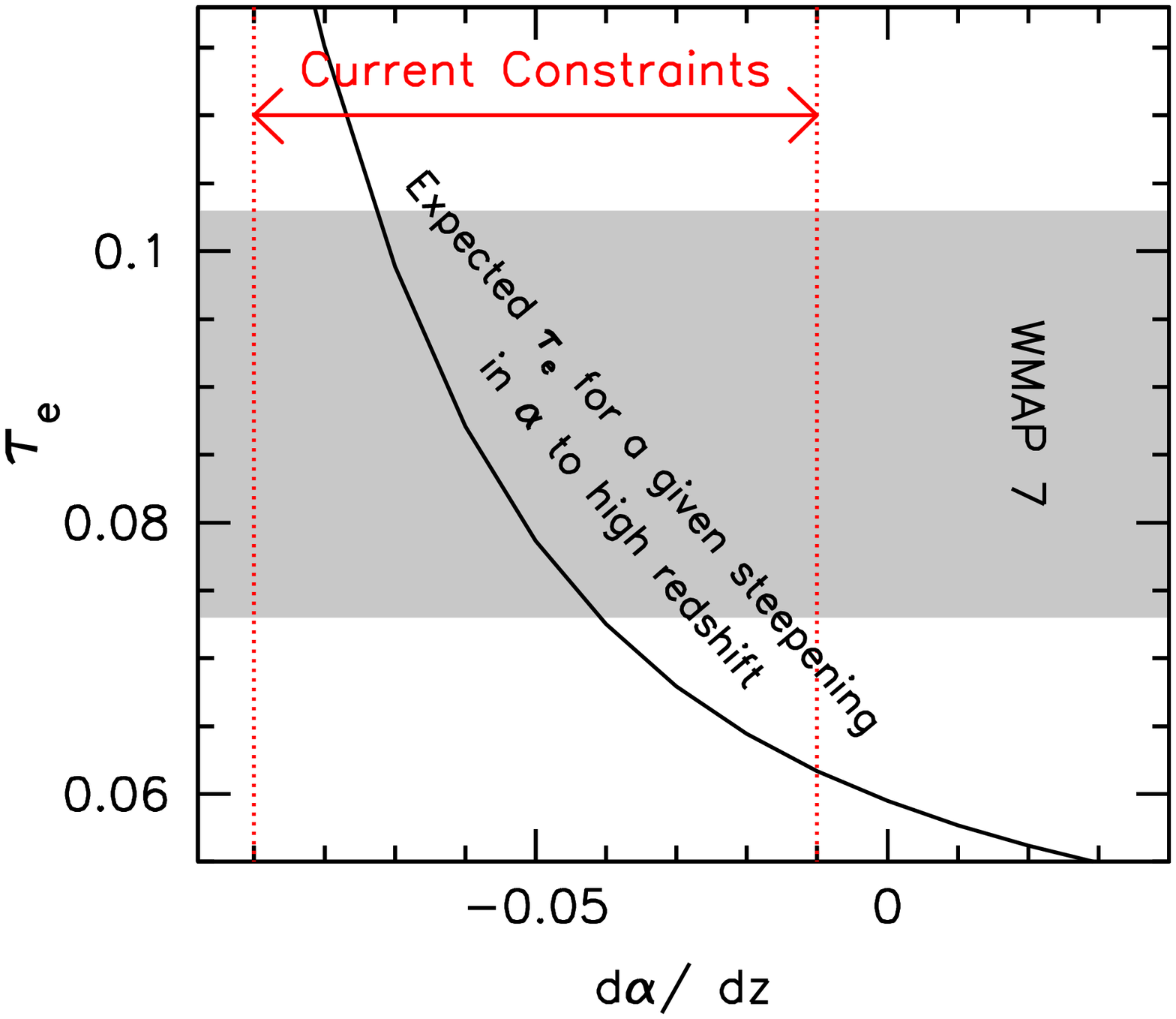}
\caption{\textit{(left)} The strong dependence of the integrated UV
  ionizing flux on the faint-end slope $\alpha$ (\S6).  The correction
  factors needed to convert the observed $UV$ photon density to the
  total density for a given faint-end slope $\alpha$ (integrating to
  the expected theoretical cut-off in the LF at $-$10$\,$mag) are
  shown for the current best estimate of the slope $\alpha$ at
  $z\sim7$ and for the upper and lower $\pm$1$\sigma$ limits of
  $\alpha\sim-1.8$ and $\alpha\sim-2.2$.  The current uncertainties in
  the correction factor are large (the ratio between the upper and
  lower 1$\sigma$ correction factors is a factor of $\sim$6).  With
  even deeper $\sim$30$\,$mag data over $\sim$10 arcmin$^2$ (two
  WFC3/IR fields), it would be possible to reduce the uncertainties in
  the correction factor to 1.8 (for the expected 1$\sigma$ error in
  $\alpha$ of $\pm$0.08).  \textit{(right)} The Thomson optical depth
  versus the change in the faint-end slope $\alpha$ per unit redshift
  (\textit{black line}).  The Thomson optical depths inferred from the
  WMAP seven-year measurements are shown in cyan.  Even a mild
  evolution in the faint-end slope $\alpha$ to higher redshift can
  have a dramatic impact on the overall output of ionizing photons
  from low luminosity galaxies.  Our current constraint on this
  evolution \textit{(red arrows)} is very poor, but suggests that
  lower luminosity galaxies may be capable of reionizing the universe.
  The evolution of the faint-end slope $\alpha$ could be significantly
  improved (over the redshift range $z\sim5$-8) through much deeper
  observations at both optical and near-IR
  wavelengths.\label{fig:prop}}
\end{figure*}

\section{The Faint-end Slope of the $UV$ LF at $z\geq6$ and Best-Fit Evolution}

These recent, new luminosity functions at z$\sim$5, 6, 7 and 8, when
combined with the existing deep z$\sim$4 LF (Bouwens et al.\ 2007),
provide our best opportunity yet to constrain the evolution of their
faint-end slopes $\alpha$ with redshift.

Figure~\ref{fig:alpha} shows the derived values of the faint-end slope
$\alpha$ versus redshift.  While the uncertainties are large, the
faint-end slopes at $z\geq6$ appear to be slightly steeper than at
lower redshifts.  A simple averaging of our results at
$z\sim6$,$~$7,$~$8 yields $-$1.87$\pm$0.13.  This is the best estimate
of the faint-end slope that we have in the reionization epoch.
However, we can also ask whether the data suggest some trend with
cosmic time.  Using $z\geq4$ LFs, we find the following best-fit trend
in $\alpha$ with redshift (taking into account the multi-variate
$\chi^2$ distribution):
\begin{equation}
\alpha = (-1.84\pm0.05) - (0.05\pm0.04) (z - 6)
\end{equation}
(see also Table~1).  The faint-end slope is steeper at earlier times,
but this of course is only of marginal significance ($1.3\sigma$).
Steepening of the faint-end slope $\alpha$ would be consistent with
various theoretical model predictions (e.g., Trenti et al.\ 2010;
Salvaterra et al.\ 2011; Jaacks et al.\ 2012).  While current evidence
for evolution in $\alpha$ is weak, this is an important first step in
quantifying this evolution and assessing the impact of an evolving
$\alpha$ on reionization.

\section{Implications for Reionization}

The very steep faint-end slopes determined for the $UV$ LFs at
$z\sim5$-8 have important implications for the role galaxies might
play in the reionization of the universe.  To investigate this, we
consider an evolving LF with the faint-end slope $\alpha$ equal to
mean value $-1.87\pm0.13$ found at $z\sim6$-8 (Figure~\ref{fig:lf} and
\ref{fig:alpha}).  The values of the $M^*$ and $\phi^*$ for our
evolving LF are taken from our empirical fitting formula to the
$z\sim4$-8 LF results (see Table~1: derived as in Bouwens et
al.\ 2008).

We compute the time evolution of the filling factor of ionized
hydrogen $Q_{HII}$ using the following relation we adapted from Madau
et al.\ (1999):
\begin{equation}\frac{dQ_{HII}}{dt}=\frac{-Q_{HII}}{t_{rec}}+\frac{\rho(SFR)_{uncorr}(z)f_{esc}10^{53.2}\textrm{photon}\,\textrm{s}^{-1}}{n_H(0)}\label{eq:xe}\end{equation}
where $f_{esc}$ is the escape fraction of Lyman-continuum photons into
the IGM, $n_H$ corresponds to the comoving volume density of neutral
hydrogen in the universe, $t_{rec}$ corresponds to the recombination
time for neutral hydrogen, and $\rho(SFR)_{uncorr} (z)$ is the SFR
density uncorrected for dust extinction.  In deriving the SFR density,
we integrate the LF down to $-10\,$mag, given the likely supression of
galaxy formation at smaller scales from the UV background, SNe
feedback, and inefficient gas cooling (e.g., Read et al.\ 2006;
Dijkstra et al.\ 2004).

To account for the increased ionizing efficiency (by up to 30\%) of
low metallicity stars expected to make up galaxies in the early
universe, we assume $10^{53.2}$ photons s$^{-1}$ per
$M_{\odot}\,$yr$^{-1}$ (Schaerer 2003).  We take $f_{esc}$ to be
$\sim$20\% motivated by the observations of Shapley et al.\ (2006) and
Iwata et al.\ (2008), but acknowledge that $f_{esc}$ is still very
poorly determined at $z\sim2$-3 and has not been determined at all at
$z\geq4$.  The recombination time $t_{rec}$ is taken to equal
\begin{equation}t_{rec}=1.0\textrm{Gyr}\left(\frac{1+z}{7}\right)^{-3}C_{3}^{-1}\label{eq:recomb}\end{equation}
where $C_{3}$ is the clumping factor of neutral hydrogen
$(<\rho>^2/<\rho^2>)$/3.  Motivated by Bolton \& Haehnelt (2007) and
Pawlik et al.\ (2009), we adopt a clumping factor of 3.

These equations were derived from Eq.~(11), Eq.~(27), and Eq.~(28) of
Madau et al.\ (1999).  We adopt only one free electron as appropriate
for single-ionized helium (e.g., Chary 2008) and take $\Omega_b
h^2=0.02260\pm0.00053$ (Komatsu et al.\ 2011).  The above equation
also assumes an IGM temperature of $\sim2\times10^{4}\,$K around the
time of reionization instead of the $\sim10^4\,$K assumed by Madau et
al.\ (1999).  This is to account for the substantial heating of the
IGM due to the reionization process itself (Hui \& Haiman 2003).  The
higher temperatures effectively double the time it takes atomic
hydrogen to recombine (Stiavelli et al.\ 2004).

With these assumptions, our mean faint-end slope $\alpha=-1.87\pm0.13$
at $z\sim6$-8, and parametrized evolution for $M^*$ and $\phi^*$ at
$z\geq4$ (Table~1), we calculate the Thomson optical depth $\tau_e$
and the filling factor of ionized hydrogen $Q_{HII}$ versus redshift.
These are shown in Figure~3.  The filling factor of ionized hydrogen
$Q_{HII}$ approaches unity at z$\sim$7 and the Thomson optical depth
is $0.061_{-0.006}^{+0.009}$.  This is short of the $\tau=0.088\pm0.015$
measured for the WMAP seven-year data set (Komatsu et al.\ 2011).

Our strongest result is based on the best estimate of the slope at
$z\sim6$-8.  Of course, if the faint-end slope $\alpha$ steepens at
earlier times (as has been suggested by theory: Trenti et al.\ 2010),
the impact of galaxies would be more substantial.  Assuming that
$\alpha$ follows the best-fit evolution with redshift (Eq.~1), we
repeat the calculations shown in Figure~3; we find that the universe
is reionized by $z\sim8$ and the Thomson optical depth is
$0.079_{-0.017}^{+0.063}$ (propagating the uncertainties on $\alpha$).
The derived optical depth extends to much larger values than using a
fixed $\alpha=-1.87\pm0.13$ and is easily consistent with the values
measured in the seven-year WMAP data.  However the large uncertainties
in the derived optical depths here point to the need to quantify the
trend in $\alpha$ vs. redshift, $d\alpha/dz$, at much greater
precision.

The present estimates of the Thomson optical depths and contribution
of galaxies to reionization are higher than in many purely
observationally-based discussions (e.g., Bouwens et al.\ 2011a; Stark
et al. 2007; Chary 2008; Oesch et al.  2009; Ouchi et al. 2009; Pawlik
et al.  2009; Bunker et al.\ 2010; Labb{\'e} et al. 2010; Robertson et
al. 2010b).  This is a direct result of the very steep faint-end
slopes $\alpha$ we find at $z\sim6$-8 and our allowing for a
steepening of these slopes to higher redshifts (based on the observed,
albeit uncertain trend).

The rapid increase in the numbers of galaxies at faint-end slopes
$\alpha$ of $-$2 is very important.  At such steep slopes virtually
all the UV flux from galaxies is output by very low luminosity
galaxies below $-$16 AB mag, even with a cutoff at $-10\,$mag.  The
adopted limit at $-10\,$mag is well-motivated by theoretical
expectations (e.g., Read et al.\ 2006) but remains a key assumption
(see Mu{\~n}oz \& Loeb 2011).

Figure~\ref{fig:prop}a illustrates how sensitive the ionizing flux
densities are to the faint-end slope (integrated to a faint-end limit
of $-10\,$mag).  Faint-end slopes of $-$2.2, $-$2.0, $-$1.8 produce
7$\times$, 2.5$\times$, and 1.5$\times$ larger luminosity densities,
respectively, than do slopes of $-$1.7 (observed at $z\sim4$.)
Changing the faint-end slope $\alpha$ by just 0.2 at $z\geq6$
(relative to our evolving faint-end slope scenario) yields Thomson
optical depths $\tau$ ranging from 0.059 to 0.107.  Changes in the
adopted cosmology will also affect the derived $\tau$; the standard
cosmology adopted here results in somewhat lower values of $\tau$ than
the WMAP7 cosmology.

One caveat in the above discussion is that our LF-fitting formula
overproduces the number of bright galaxies found in recent $z\sim10$
searches (Bouwens et al.\ 2011a; Oesch et al.\ 2011).  Our fitting
formula would then overpredict the total ionizing UV flux from $z>8$
galaxies.  For example, a 20\% decrease in the ionizing flux from
$z>8$ galaxies (at the bright end of the LF) would lower the Thomson
optical depths by 0.005.

Of course, other ingredients may also play an important role in
generating the high Thomson optical depths measured by WMAP.  These
include an even smaller clumping factor (e.g., Bolton \& Haehnelt
2007; Pawlik et al. 2009) and perhaps a contribution from early
population III stars to the UV photon flux (e.g., Cen 2003; Ricotti \&
Ostriker 2004).

\section{Implications for Future Observations}

Can these new and intriguing results be improved?  Key issues are how
well the faint-end slope $\alpha$ can be determined at $z>6$ and what
constraints (most-likely theoretical) can be placed on the
low-luminosity cutoff of the galaxy LF.  As shown in
Figures~\ref{fig:reion}-\ref{fig:prop}, both the Thomson optical
depths and the filling factors for ionized hydrogen inferred from
recent measurements of the $z\geq6$ LFs are \textit{very} sensitive to
their faint-end slopes.

The current uncertainty in the integrated luminosity density is 0.8
dex, i.e., a factor of 6 (Figure~4a).  To determine the total
luminosity density within 0.3 dex, the faint-end slope needs to be
constrained to $<$0.1.  Fortunately, this is now practical with the
new WFC3/IR camera on HST, by extending current observations to 30 mag
over 10$\,$arcmin$^2$ and leveraging deep wide-area data (e.g., from
CANDELS and BORG/HIPPIES: Trenti et al.\ 2011).

To provide robustness to the results, measurements of the faint-end
slope will be necessary at several different redshifts, i.e., at
$z\sim5$, $z\sim6$, $z\sim7$, and $z\sim8$, and not simply at
$z\sim7$-8.  By making this measurement at multiple redshifts,
$\alpha$ can be determined more precisely.  Furthermore, the
\textit{evolution} of $\alpha$ with cosmic time (Figures~2, 4) allows
for a plausible extrapolation to $z>8$ where the uncertainties from
direct measurement will remain large.

Deep WFC3/IR data from the HUDF09 program indicate that
lower-luminosity galaxies are likely to be the primary source of UV
photons needed to reionize the universe.  More definitive results on
this important and long-standing problem are now within reach with
HST.

\acknowledgements

Comments by our referee significantly improved this paper.  We
acknowledge support from NASA grant NAG5-7697, NASA grant
HST-GO-11563, and ERC grant HIGHZ \#227749.  PO acknowledges support
from NASA through a Hubble Fellowship grant \#51278.01 awarded by
STScI.

\begin{deluxetable*}{ccccccccc}
\tabletypesize{\scriptsize}
\tablecaption{Observational Data, High-Redshift Samples, and Best-Fit Schechter Parameters\label{tab}}
\tablewidth{17cm}
\startdata
 \tableline \\
 \multicolumn{9}{c}{Observational Data} \\
 \tableline
\colhead{} & \colhead{Area} & \multicolumn{7}{c}{Depth ($5\sigma$)} \\
\colhead{Field} & \colhead{(arcmin$^2$)} & \colhead{$B_{435}$}
& \colhead{$V_{606}$} & \colhead{$i_{775}$} & \colhead{$z_{850}$}
& \colhead{$Y_{105}$} & \colhead{$J_{125}$} & \colhead{$H_{160}$} \\
 \tableline
HUDF09 & 4.7 & 29.7 & 30.1 & 29.9 & 29.4 & 29.6 & 29.9 & 29.9 \\
HUDF09-1 & 4.7 & --- & 29.0 & 29.0 & 29.0 & 29.0 & 29.3 & 29.1 \\
HUDF09-2 & 4.7 & 28.8 & 29.9 & 29.3 & 29.2 & 29.2 & 29.5 & 29.3 \\
ERS & 39.2 & 28.2 & 28.5 & 28.0 & 28.0 & 27.9 & 28.4 & 28.1 \\
CANDELS-DEEP & 63.1 & 28.2 & 28.5 & 28.0 & 28.0 & --- & 28.1 & 27.8 \\
CANDELS-WIDE & 41.9 & 28.2 & 28.5 & 28.0 & 28.0 & --- & 27.8 & 27.5 \\
\tableline \\
 \multicolumn{9}{c}{High-Redshift Samples} \\
 \tableline
\colhead{} & \multicolumn{2}{c}{Ultra-Deep (HUDF09)} & \multicolumn{2}{c}{Ultra-Deep (HUDF09-1)} & \multicolumn{2}{c}{Ultra-Deep (HUDF09-2)} & \multicolumn{2}{c}{Wide (ERS+CANDELS)}\\
\colhead{Redshift} & \colhead{\#} & \colhead{Limit} & \colhead{\#} & \colhead{Limit} & \colhead{\#} & \colhead{Limit} & \colhead{\#} & \colhead{Limit}\\
 \tableline
$z\sim5$ & 57 & $J\leq29.7$ & 85 & $J\leq29.2$ & 101 & $J\leq29.2$ & 264 & $J\leq28$ \\
$z\sim6$ & 56 & $J\leq29.7$ & 27 & $J\leq29.2$ & 23 & $J\leq29.2$ & 97 & $J\leq28$ \\
$z\sim7$ & 29 & $J\leq29.5$ & 17 & $J\leq29$ & 14 & $J\leq29$ & 13 & $J\leq28$ \\
$z\sim8$ & 24 & $H\leq29.5$ & 14 & $J\leq29$ & 15 & $H\leq29$ & 6 & $H\leq27.5$ \\
\tableline \\
 \multicolumn{5}{c}{Best-Fit Schechter Parameters} \\
 \tableline
\colhead{Redshift} & \colhead{$M_{UV,AB}^{*}$} & \colhead{$\phi^*$} & \colhead{$\alpha$} & \colhead{Reference\tablenotemark{a}} & \multicolumn{4}{c}{LF Fitting Formula at $z\geq4$} \\
\tableline
$z\sim4$ & $-$20.98$\pm$0.10 & 1.3$\pm$0.2 & $-$1.73$\pm$0.05 & [1] & \multicolumn{4}{c}{$M_{UV,AB} ^{*} = (-20.34\pm0.11) + (0.28\pm0.06) (z-6)$} \\
$z\sim5$ & $-$20.60$\pm$0.23 & 1.4$_{-0.5}^{+0.7}$ & $-$1.79$\pm$0.12 & This Work & \multicolumn{4}{c}{$\phi ^{*} = 10^{-2.90\pm 0.09 + (-0.04\pm0.05) (z-6)}$} \\
$z\sim6$ & $-$20.37$\pm$0.30 & 1.4$_{-0.6}^{+1.1}$ & $-$1.73$\pm$0.20 & This Work & \multicolumn{4}{c}{$\alpha = (-1.84\pm0.05) - (0.05\pm0.04) (z - 6)$} \\ 
$z\sim7$ & $-$20.14$\pm$0.26 & 0.86$_{-0.39}^{+0.70}$ & $-$2.01$\pm$0.21 & [2] \\
$z\sim8$ & $-$20.10$\pm$0.52 & 0.59$_{-0.37}^{+1.01}$ & $-$1.91$\pm$0.32 & [2]
\enddata
\tablenotetext{a}{References: [1] Bouwens et al. 2007, [2] Bouwens et al. 2011}
\end{deluxetable*}


\begin{thebibliography}{} 
\bibitem[Beckwith et al.(2006)]{2006AJ....132.1729B} Beckwith, S.~V.~W., et 
al.\ 2006, \aj, 132, 1729
\bibitem[Bolton \& Haehnelt(2007)]{2007MNRAS.382..325B} Bolton, J.~S.,
  \& Haehnelt, M.~G.\ 2007, \mnras, 382, 325
\bibitem[Bouwens et al.(2007)]{2007ApJ...670..928B} Bouwens, R.~J., 
Illingworth, G.~D., Franx, M., \& Ford, H.\ 2007, \apj, 670, 928
\bibitem[Bouwens et al.\ (2008)]{2008ApJ...686..230B} Bouwens, R.~J., 
Illingworth, G.~D., Franx, M., \& Ford, H.\ 2008, \apj, 686, 230 
\bibitem[Bouwens et al. (2010b)]{eee2} Bouwens, R.J., et al.\ 2010b,
  \apj, 709, L133
\bibitem[Bouwens et al.(2011)]{2011Natur.469..504B} Bouwens, R.~J., et al.\ 
2011a, \nat, 469, 504 
\bibitem[Bouwens et al.(2011)]{2011ApJ...737...90B} Bouwens, R.~J., 
Illingworth, G.~D., Oesch, P.~A., et al.\ 2011b, \apj, 737, 90
\bibitem[Bouwens et al.(2011)]{2011arXiv1109.0994B} Bouwens, R.~J.,
  Illingworth, G.~D., Oesch, P.~A., et al.\ 2011c, \apj, submitted,
  arXiv:1109.0994
\bibitem[Bunker et al.(2010)]{2010MNRAS.409..855B} Bunker, A.~J., et al.\ 
2010, \mnras, 409, 855
\bibitem[Cen(2003)]{2003ApJ...591...12C} Cen, R.\ 2003, \apj, 591, 12
\bibitem[Chary(2008)]{2008ApJ...680...32C} Chary, R.-R.\ 2008, \apj, 680, 
32 
\bibitem[Dijkstra et al.(2004)]{2004ApJ...601..666D} Dijkstra, M., Haiman, 
Z., Rees, M.~J., \& Weinberg, D.~H.\ 2004, \apj, 601, 666
\bibitem[Efstathiou et al.(1988)]{1988MNRAS.232..431E} Efstathiou, G., 
Ellis, R.~S., \& Peterson, B.~A.\ 1988, \mnras, 232, 431
\bibitem[Fan et al.(2002)]{2002AJ....123.1247F} Fan, X., Narayanan, V.~K., 
Strauss, M.~A., White, R.~L., Becker, R.~H., Pentericci, L., 
\& Rix, H.-W.\ 2002, \aj, 123, 1247
\bibitem[Grogin et al.(2011)]{2011ApJS..197...35G} Grogin, N.~A., Kocevski, 
D.~D., Faber, S.~M., et al.\ 2011, \apjs, 197, 35 
\bibitem[Hathi et al.(2010)]{2010ApJ...720.1708H} Hathi, N.~P., et al.\ 
2010, \apj, 720, 1708
\bibitem[Hui \& Haiman(2003)]{2003ApJ...596....9H} Hui, L., \& Haiman, Z.\ 
2003, \apj, 596, 9
\bibitem[Iwata et al.(2009)]{2009ApJ...692.1287I} Iwata, I., et al.\ 2009, 
\apj, 692, 1287
\bibitem[Jaacks et al.(2012)]{2012MNRAS.420.1606J} Jaacks, J., Choi, J.-H., 
Nagamine, K., Thompson, R., \& Varghese, S.\ 2012, \mnras, 420, 1606 
\bibitem[Koekemoer et al.(2011)]{2011ApJS..197...36K} Koekemoer, A.~M., 
Faber, S.~M., Ferguson, H.~C., et al.\ 2011, \apjs, 197, 36 
\bibitem[Komatsu et al.(2011)]{2011ApJS..192...18K} Komatsu, E., et al.\ 
2011, \apjs, 192, 18 
\bibitem[Labbe et al.(2010)]{2009arXiv0911.1356L} Labb{\'e}, I., et
  al.\ 2010, \apjl, 716, L103
\bibitem[Madau, Haardt, \& Rees(1999)]{1999ApJ...514..648M} Madau, P.,
Haardt, F., \& Rees, M.~J.\ 1999, \apj, 514, 648
\bibitem[Mu{\~n}oz 
\& Loeb(2011)]{2011ApJ...729...99M} Mu{\~n}oz, J.~A., \& Loeb, A.\ 2011, \apj, 729, 99 
\bibitem[Oesch et al.(2007)]{2007ApJ...671.1212O} Oesch, P.~A., et al.\ 
2007, \apj, 671, 1212 
\bibitem[Oesch et al.(2009)]{2009ApJ...690.1350O} Oesch, P.~A., et al.\ 
2009, \apj, 690, 1350
\bibitem[Oesch et al. (2010)]{eee} Oesch, P.A., et al.\ 2010a, \apj,
709, L16
\bibitem[Oesch et al. (2010b)]{eeef} Oesch, P.A., et al.\ 2010b, \apj,
709, L21
\bibitem[Oesch et al.(2010)]{2010ApJ...725L.150O} Oesch, P.~A., et al.\ 
2010c, \apjl, 725, L150
\bibitem[Oesch et al.(2012)]{2012ApJ...745..110O} Oesch, P.~A., Bouwens, 
R.~J., Illingworth, G.~D., et al.\ 2012a, \apj, 745, 110
\bibitem[Oesch et al.(2012)]{2012arXiv1201.0755O} Oesch, P.~A., Bouwens, 
R.~J., Illingworth, G.~D., et al.\ 2012b, \apj, submitted, arXiv:1201.0755 
\bibitem[Oke \& Gunn(1983)]{1983ApJ...266..713O} Oke, J.~B., \& Gunn, 
J.~E.\ 1983, \apj, 266, 713 
\bibitem[Ouchi et al.(2009)]{2009ApJ...706.1136O} Ouchi, M., et al.\ 2009, 
\apj, 706, 1136
\bibitem[Ouchi et al.(2010)]{2010ApJ...723..869O} Ouchi, M., et al.\ 2010, 
\apj, 723, 869 
\bibitem[Pawlik et al.(2009)]{2009MNRAS.394.1812P} Pawlik, A.~H.,
  Schaye, J., \& van Scherpenzeel, E.\ 2009, \mnras, 394, 1812
\bibitem[Read et al.(2006)]{2006MNRAS.371..885R} Read, J.~I., Pontzen, 
A.~P., \& Viel, M.\ 2006, \mnras, 371, 885 
\bibitem[Reddy \& Steidel(2009)]{2009ApJ...692..778R} Reddy, N.~A., \&
  Steidel, C.~C.\ 2009, \apj, 692, 778
\bibitem[Ricotti 
\& Ostriker(2004)]{2004MNRAS.350..539R} Ricotti, M., \& Ostriker, J.~P.\ 2004, \mnras, 350, 539
\bibitem[Robertson(2010)]{2010ApJ...713.1266R} Robertson, B.~E.\ 2010a, 
\apj, 713, 1266 
\bibitem[Robertson et al.(2010)]{2010Natur.468...49R} Robertson, B.~E., 
Ellis, R.~S., Dunlop, J.~S., McLure, R.~J., 
\& Stark, D.~P.\ 2010b, \nat, 468, 49
\bibitem[Salvaterra et al.(2011)]{2011MNRAS.414..847S} Salvaterra, R., 
Ferrara, A., \& Dayal, P.\ 2011, \mnras, 414, 847 
\bibitem[Sandage, Tammann, \& Yahil(1979)]{1979ApJ...232..352S} Sandage, 
A., Tammann, G.~A., \& Yahil, A.\ 1979, \apj, 232, 352
\bibitem[Schaerer(2003)]{2003A&A...397..527S} Schaerer, D.\ 2003, \aap, 
397, 527 
\bibitem[Shapley et al.(2006)]{2006ApJ...651..688S} Shapley, A.~E., 
Steidel, C.~C., Pettini, M., Adelberger, K.~L., \& Erb, D.~K.\ 2006, \apj, 
651, 688 
\bibitem[Stark et al.(2007)]{2007ApJ...659...84S} Stark, D.~P., Bunker, 
A.~J., Ellis, R.~S., Eyles, L.~P., \& Lacy, M.\ 2007, \apj, 659, 84 
\bibitem[Stiavelli et al.(2004)]{2004ApJ...610L...1S} Stiavelli, M., Fall, 
S.~M., \& Panagia, N.\ 2004, \apjl, 610, L1
\bibitem[Su et al.(2011)]{2011ApJ...738..123S} Su, J., Stiavelli, M., 
Oesch, P., et al.\ 2011, \apj, 738, 123
\bibitem[Trenti \& Stiavelli(2008)]{2008ApJ...676..767T} Trenti, M.,
  \& Stiavelli, M.\ 2008, \apj, 676, 767
\bibitem[Trenti et al.(2010)]{2010ApJ...714L.202T} Trenti, M.,
  Stiavelli, M., Bouwens, R.~J., Oesch, P., Shull, J.~M., Illingworth,
  G.~D., Bradley, L.~D., \& Carollo, C.~M.\ 2010, \apjl, 714, L202
\bibitem[Trenti et al.(2011)]{2011ApJ...727L..39T} Trenti, M., Bradley, 
L.~D., Stiavelli, M., et al.\ 2011, \apjl, 727, L39
\bibitem[Willott et al.(2010)]{2010AJ....139..906W} Willott, C.~J., 
Delorme, P., Reyl{\'e}, C., et al.\ 2010, \aj, 139, 906
\bibitem[Windhorst et al.(2011)]{2011ApJS..193...27W} Windhorst, R.~A., et 
al.\ 2011, \apjs, 193, 27
\end{thebibliography}
\end{document}